# Cognitive Radio Simultaneous Spectrum Access/ One-shot Game Modelling

Ligia C. Cremene[*,***], D. Dumitrescu[**,***], Rėka Nagy[**], Noemi Gasko[**]
[*] Technical University of Cluj-Napoca, Department of Communications, Adaptive Systems Laboratory, Romania
[**] Babeş-Bolyai University, Department of Computer Science, Centre for the Study of Complexity, Cluj-Napoca, Romania
[***] Romanian Institute of Science and Technology, Cluj-Napoca, Romania
ligia.cremene@com.utcluj.ro, {ddumitr, reka, gasko}@cs.ubbcluj.ro

*Abstract*— **The aim of this paper is to asses simultaneous spectrum access situations that may occur in Cognitive Radio (CR) environments. The approach is that of one-shot, non-cooperative games describing CR interactions. Open spectrum access scenarios are modelled based on continuous and discrete reformulations of the Cournot game theoretical model. CR interaction situations are described by Nash and Pareto equilibria. Also, the heterogeneity of players is captured by the new concept of joint Nash-Pareto equilibrium, allowing CRs to be biased toward different types of equilibrium. Numerical simulations reveal equilibrium situations that may be reached in simultaneous access scenarios of two and three users.**

## I. INTRODUCTION

Cognitive radio (CR) technology is seen as the key enabler for next generation communication networks, which will be spectrum-aware, dynamic spectrum access networks [1], [2], [3]. Cognitive radios (CRs) hold the promise for an efficient use of the radio resources and are seen as the solution to the current low usage of the radio spectrum [2], [4], [5]. In a CR environment users strategically compete for spectrum resources in dynamic scenarios.

In this paper the problem of simultaneous, open spectrum access is addressed from a game theoretical perspective. Game Theory (GT) provides a fertile framework and the computational tools for CR interaction analysis. By devising GT simulations, insight may be gained on unanticipated situations that may arise in spectrum access. Clearly CR interactions are strategic interactions [8]: the utility of one CR agent/player depends on the actions of all the other CRs in the area.

The proposed approach relies on the following assumptions: *(i)* CRs have perfect channel sensing and RF reconfiguration capabilities [2], [6], [7], *(ii)* CRs are myopic, self-regarding players, *(iii)* repeated interaction among the same CRs is not likely to occur on a regular basis [9], and *(iv)* CRs do not know in advance what actions the other CRs will choose.

These are reasons to consider one-shot, non-cooperative games for the open spectrum access analysis.

An oligopoly competition game model – Cournot – is reformulated in terms of spectrum access. Continuous and discrete instances of the game are analyzed.

Nash and Pareto equilibria are revisited for the discrete instance of the game. Heterogeneity of players is captured by joint Nash-Pareto equilibria, allowing CRs to be biased toward different types of equilibrium.

The paper is structured as follows: Section II provides some basic insights on game-equilibria detection. The reformulation of Cournot game theoretic model for simultaneous, open spectrum access is described in Section III. Section IV discusses simulation results obtained for continuous and discrete instances of the game. Conclusions are presented in Section V.

## II. GAME EQUILIBRIA IN BRIEF

A strategic-form game model is defined by its three major components: a finite set of players, a set of actions, and a payoff/utility function which measures the outcome for each player, determined by the actions of all players [8], [10].

A game may be defined as a system $G = ((N, S_i, u_i), i = 1,\ldots, n)$ where:

(i) $N$ represents the set of $n$ players, $N = \{1,\ldots, n\}$.

(ii) for each player $i \in N$, $S_i$ represents the set of actions $S_i = \{s_{i1}, s_{i2}, \ldots, s_{im}\}$; $S = S_1 \times S_2 \times \ldots S_n$ is the set of all possible game situations;

(iii) for each player $i \in N$, $u_i : S \rightarrow R$ represents the payoff function.

A strategy profile is a vector $s = (s_1,\ldots, s_n)$, where $s_i \in S_i$ is a strategy (or action) of player $i$. By $(s_i, s^*_{-i})$ we denote the strategy profile obtained from $s^*$ by replacing the strategy of player $i$ with $s_i$, i.e. $(s_i, s^*_{-i}) = (s^*_1, s^*_2,\ldots, s^*_{i-1}, s_i, s^*_{i+1},\ldots, s^*_n)$.

A strategy profile in which each player's strategy is a best response to the strategies of the other players is a *Nash equilibrium* (NE) [8], [11]. Informally, a strategy profile is a Nash equilibrium if no player can improve her payoff by unilateral deviation.

Considering two strategy profiles $x$ and $y$ from $S$, the strategy profile $x$ is said to Pareto dominate the strategy profile

$y$ (and we write $x <_P y$) if the payoff of each player using strategy $x$ is greater or equal to the payoff associated to strategy $y$ and at least one payoff is strictly greater. The set of all non-dominated strategies (Pareto frontier) represents the set of *Pareto equilibria* of the game [8].

In an $n$-player game consider that each player $i$ acts based on a certain type of rationality $r_i$, $i = 1,\ldots, n$. We may consider a three-player game where $r_1$ = Nash, $r_2$ = Nash, and $r_3$ = Pareto. The first two players are biased towards the Nash equilibrium and the other one is Pareto-biased. Thus, a new type of equilibrium, called the joint Nash-Pareto equilibrium (N-P), may be considered [15]. The considered generalization involves heterogeneous players that are biased towards different equilibrium types or may act based on different types of rationality [15].

Games can be viewed as multiobjective optimization problems, where the payoffs of the participating players are to be maximized [15]. An appealing technique is the use of generative relations and evolutionary algorithms for detecting equilibrium strategies. The payoff of each player is treated as an objective and the generative relation induces an appropriate dominance concept, which is used for fitness assignment purpose [21].

Game equilibria may be characterized by generative relations on the set of game strategies [21]. The idea is that the non-dominated strategies with respect to the generative relation equals (or approximate) the equilibrium set.

Let us denote by $I_N$ the set of Nash-biased players and by $I_P$ the set of Pareto-biased players. We may write:

$$I_N = \{i \in \{1,..,n\} | r_i = Nash\} \text{ and}$$
$$I_P = \{i \in \{1,..,n\} | r_i = Pareto\}.$$

$E(x,y)$ measures the relative efficiency of strategy $x$ with respect to strategy $y$, and is defined by:

$$E(x,y) = card(\{i \in I_N | u_i(x_i, y_{-i}) \geq u_i(y), x_i \neq y_i\}) + card(I_P)P_i(x,y),$$

where

$$P_i(x,y) = \begin{cases} 1, & if\ x <_P y, \\ 0, & otherwise. \end{cases}$$

The N-P dominance relation $<_{NP}$, defined as $x <_{NP} y$ if

$$E(y,x) < E(x,y),$$

may be considered as the generative relation for *joint Nash-Pareto equilibria*.

In the following, both continuous and discrete instances of a game are considered. As there is no direct calculus method for discrete equilibria an heuristic method based on other principles is needed. We propose the use of a method combining the algorithmic character of a game (through generative relations) with an evolutionary technique [15]. The generative relation allows comparison of two strategies and may serve for fitness assignment purposes in an evolutionary procedure.

Numerical experiments aim the detection of pure equilibria or a combination of equilibria paralleling CRs interaction. A slight modification of NSGA2 [19], called GTNSGA2 is considered.

## III. Open Spectrum Access Modelling

The problem of open spectrum access is modelled as a non-cooperative, one-shot game. We consider the Cournot standard oligopoly competition model, reformulated in terms of radio resource access. CR simultaneous access situations are considered and modelled as one-shot games. As simultaneous spectrum access scenarios do not imply large numbers of users, two and three-player games are considered relevant. Continuous and discrete instances of the game are analyzed.

We analyze different types of game equilibria, as they describe several types of strategic interactions between cognitive agents – each CR's action directly affects the other CRs payoffs.

### *Open spectrum access model – Cournot reformulation*

In the Cournot economic competition model players are firms that simultaneously choose quantities [8].

We consider a general open spectrum access scenario that can be modelled as a reformulation of the Cournot oligopoly game [12].

Suppose there are $n$ radios attempting to access the same set of available channels, simultaneously. Each CR $i$ may decide the number $c_i$ of simultaneous channels to access. The question is how many simultaneous channels should each CR access in order to maximize its operation efficiency?

As mentioned before, a strategic-form game model is defined by its three major components: set of players, set of actions, and payoffs. For a general open access scenario the Cournot competition may be reformulated as follows:

| | |
|---|---|
| *Players* | cognitive radios simultaneously attempting to access a certain set of channels $W$; |
| *Actions* | the strategy of each player $i$ is the number $c_i$ of simultaneously accessed channels; A strategy profile is a vector $c = (c_1,\ldots,c_n)$. |
| *Payoffs* | the difference between a function of goodput $P(C)c_i$ and the cost of accessing $c_i$ simultaneous channels $Kc_i$. |

A linear inverse demand function is considered – the number of non-interfered symbols $P(C)$ is determined from the total number $C$ of accessed channels (occupied bandwidth).

The demand function may be defined as:

$$P(C) = \begin{cases} W - C, if C \leq W, \\ 0, C > W, \end{cases}$$

where $W > 0$ is the parameter of the inverse demand function, and $C = \sum_{i=1}^{n} c_i$ is the total number of accessed channels.

The goodput for CR $i$ is $P(C)c_i$ . Radio $i$'s cost for supporting $c_i$ simultaneous channels is $Kc_i$.

The payoff of CR $i$ may then be written as:

$$u_i(c) = P(C)c_i - Kc_i .$$

The payoff function is kept simple in order to focus on the emergent phenomena. The computational model allows for more complex payoff functions to be implemented, accounting for various parameters, but the essence is captured here. In general, $P$ decreases with the total number of implemented channels $C$, and the total cost for supporting $c_i$ simultaneous channels, $Kc_i$, increases with $c_i$ (more bandwidth implies more processing resources and more power consumption) [12].

If these effects are approximated by linear functions, the payoff function can be rewritten as:

$$u_i(c) = \left(W - \sum_{k=1}^{n} c_k\right)c_i - Kc_i ,$$

where
$W$ is the number of available channels, and
$K$ is the cost of accessing one channel.

The Nash equilibrium, considered the solution of this game, can be calculated as follows:

$$c_i^* = (W - K)/(n+1), \forall i \in \mathrm{N} .$$

Pareto and Nash-Pareto equilibria are described by the generative relations [15] presented in Section II.

## IV. Numerical Experiments

In order to illustrate open spectrum access situations, scenarios with two and three CRs simultaneously trying to access a given set of channels are considered. CR strategies and payoffs are represented two- and three-dimensionally. As the continuous modelling captures only partially the variety of possible equilibrium situations, discrete instances of the game are also considered.

The results represent a sub-set of more extensive simulations. For equilibria detection the evolutionary technique from [15] is considered. A population of 100 strategies is evolved using a rank based fitness assignment technique. In all experiments the process converges in less than 100 generations. Our tests indicate that the evolutionary method for equilibrium detection is scalable with respect to the number of available channels [20].

Following the game formulation in Section III, the simulation parameters are chosen: $W = 10$ (available channels) and $K = 1$ (cost of accessing one channel).

### A. Continuous Cournot modelling, 2-player simultaneous access

Simulation results are presented for the Cournot competition with two CRs simultaneously trying to access a set of channels. The stable interaction situations are captured by the detected equilibria (Fig. 1): Nash, Pareto, Nash-Pareto, and Pareto-Nash. The four types of equilibria are obtained in separate runs. Fig. 2 illustrates the payoffs of the two CRs $u_1(c_1, c_2)$ and $u_2(c_1, c_2)$.

As illustrated in Fig. 1, the NE corresponds to a situation where each of the two CRs activates 3 channels (from 10 available). The NE indicates the maximum number of channels a Nash-biased player may access without decreasing its payoff (Fig. 2). NE is a stable point from which no CR has any incentive to individually deviate.

The Pareto equilibrium (Fig. 2) describes a larger range of payoffs, capturing unbalanced as well as equitable situations (the two ends of the front vs. the middle). Although each CR tries to maximize its utility, none of them can access more than half of the available channels: $c \in [0, 4.5]$ (Fig.1). The Pareto payoffs (Fig. 2) are in the range [0, 20] and their sum is always larger than the NE payoff (9,9).

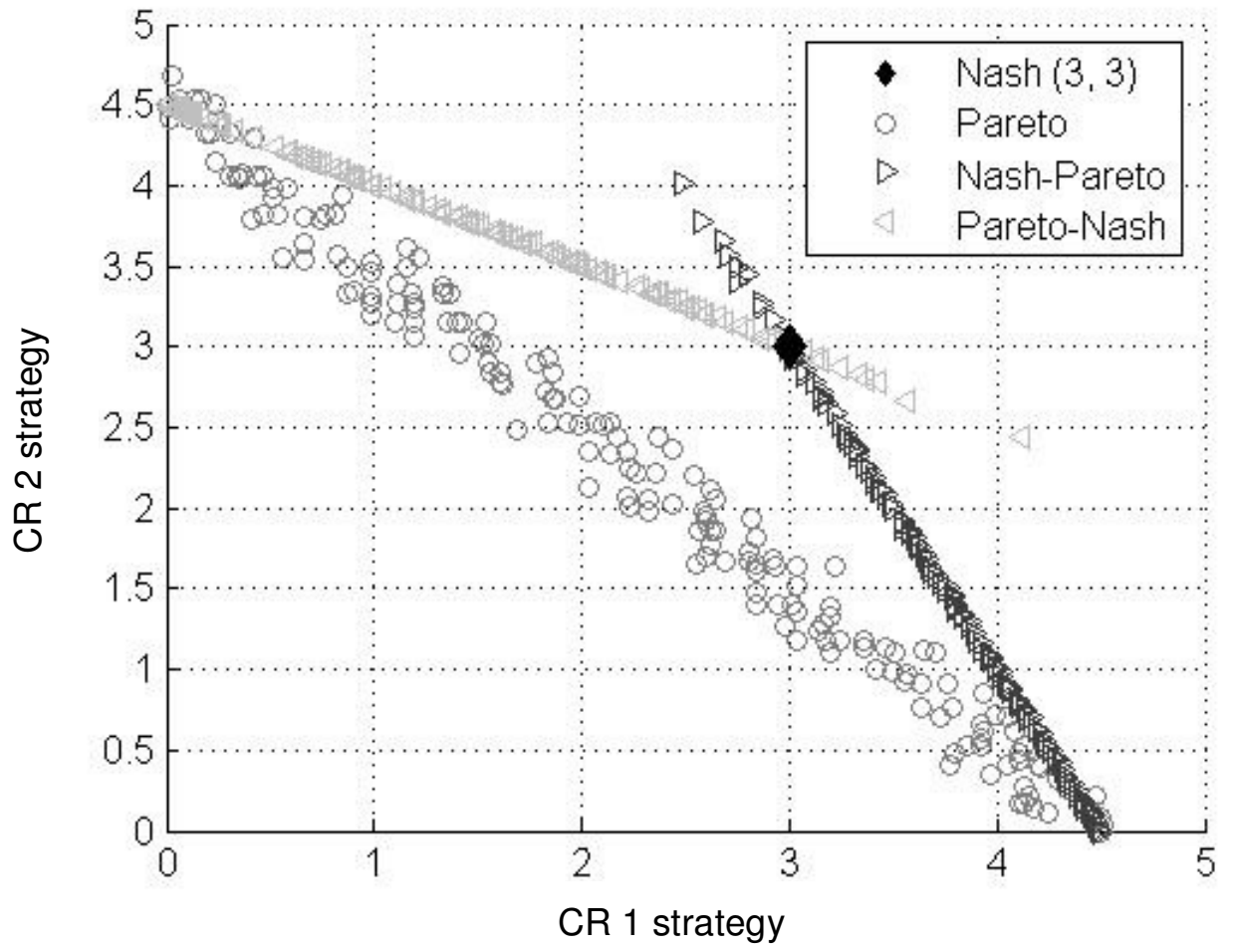

Figure 1. Cournot modelling – two radios (W = 10, K = 1). Evolutionary detected equilibria: Nash (3,3), Pareto, Nash-Pareto, and Pareto-Nash

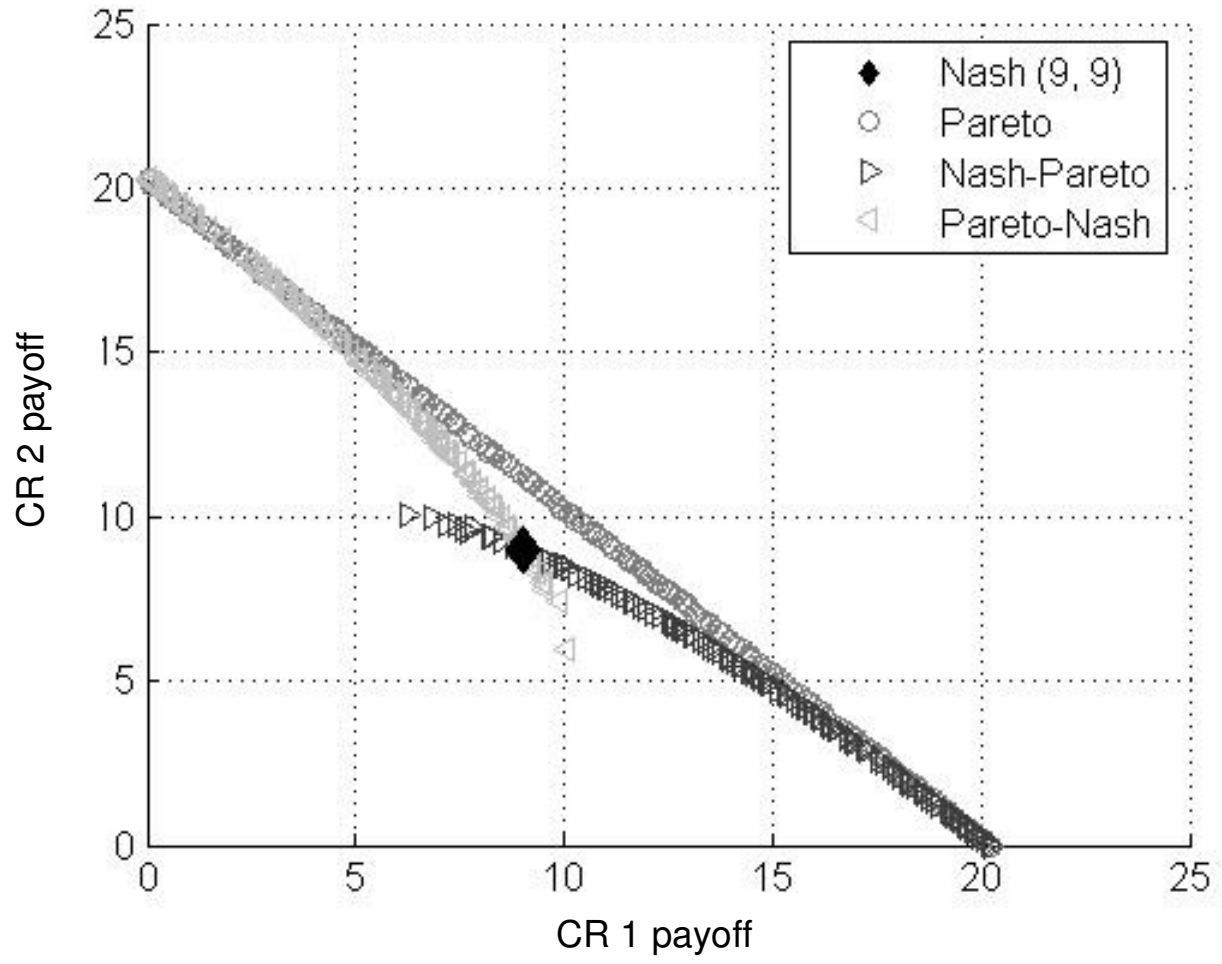

Figure 2. Cournot modelling – two radios (W = 10, K = 1). Payoffs of the evolutionary detected equilibria: Nash (9, 9), Pareto, N-P, and P-N.

In some cases, a Nash-Pareto situation enables the CR to access more channels than for the NE strategy (Fig.1). A N-P equilibrium captures the situation where one CR wants to keep its payoff (and chooses a Nash leading strategy) whereas the other CR goes for the maximum payoff (and chooses a Pareto leading strategy). In the performed experiments the P-N equilibrium is symmetric to the N-P equilibrium with respect to NE. In the continuous instance of the game none of the N-P strategies actually reach NE (Fig. 1).

Even if the N-P strategies allow the CRs to access more channels (Fig. 1), the payoffs are smaller than for the Pareto strategies (Fig. 2). This mirrors the effect of interference increasing with the number of accessed channels.

*B. Discrete Cournot modelling – 2-player simultaneous access*

Continuous Cournot modelling does not capture the variety of possible situations. A discrete instance of the game seems more realistic as CRs' strategies represent the number of accessed channels. A generalization of Cournot game allowing discrete strategies is proposed.

The evolutionary detected Nash, Pareto, N-P, and P-N equilibria, for the discrete instance of the game, are captured in Fig. 3. The discrete instance of the game reveals new equilibria: there are three NE strategies (2,4), (3,3), and (4,2). The existence of multiple NEa indicates a flexibility in choosing the number of accessed channels. There are more situations from which the CRs have no incentive to unilaterally deviate. Also, the three NE payoffs (6,12), (9,9), (12,6) offer a satisfactory diversity of utilities (Fig.4).

Whether in the continuous instance of the game none of the N-P or P-N strategies actually reach NE, in the discrete instance they overlap the three NEa (Fig. 3).

The (3,3) NE is the most stable game situation as it maintains even for N-P and P-N strategies (the overlapping of symbols in Fig.3). The other two NEa, (2,4) and (4,2), are also stable and are maintained for one of the joint strategies – N-P or P-N, respectively.

We may also notice the overlapping of most N-P and P-N equilibria onto Pareto equilibria (Fig.3 and Fig.4). This may indicate that Pareto optimality is maintained in most cases even if a CR plays Nash and the other one plays Pareto.

We may say that, for this particular instance of the game (W=10, K=1), heterogeneity of players does not affect the game equilibria.

*C. Continuous Cournot modelling, 3-player simultaneous access*

Fig. 5 and Fig. 6 illustrate the equilibrium situations for three CRs simultaneously sharing a set of channels *W=10.*

The NE (Fig. 5) corresponds to a situation where each of the three CRs activates 2 channels (from 10 available). The NE indicates the maximum number of channels a Nash-biased CR may access without decreasing its payoff (Fig. 5). NE is a stable point from which no CR has any incentive to individually deviate.

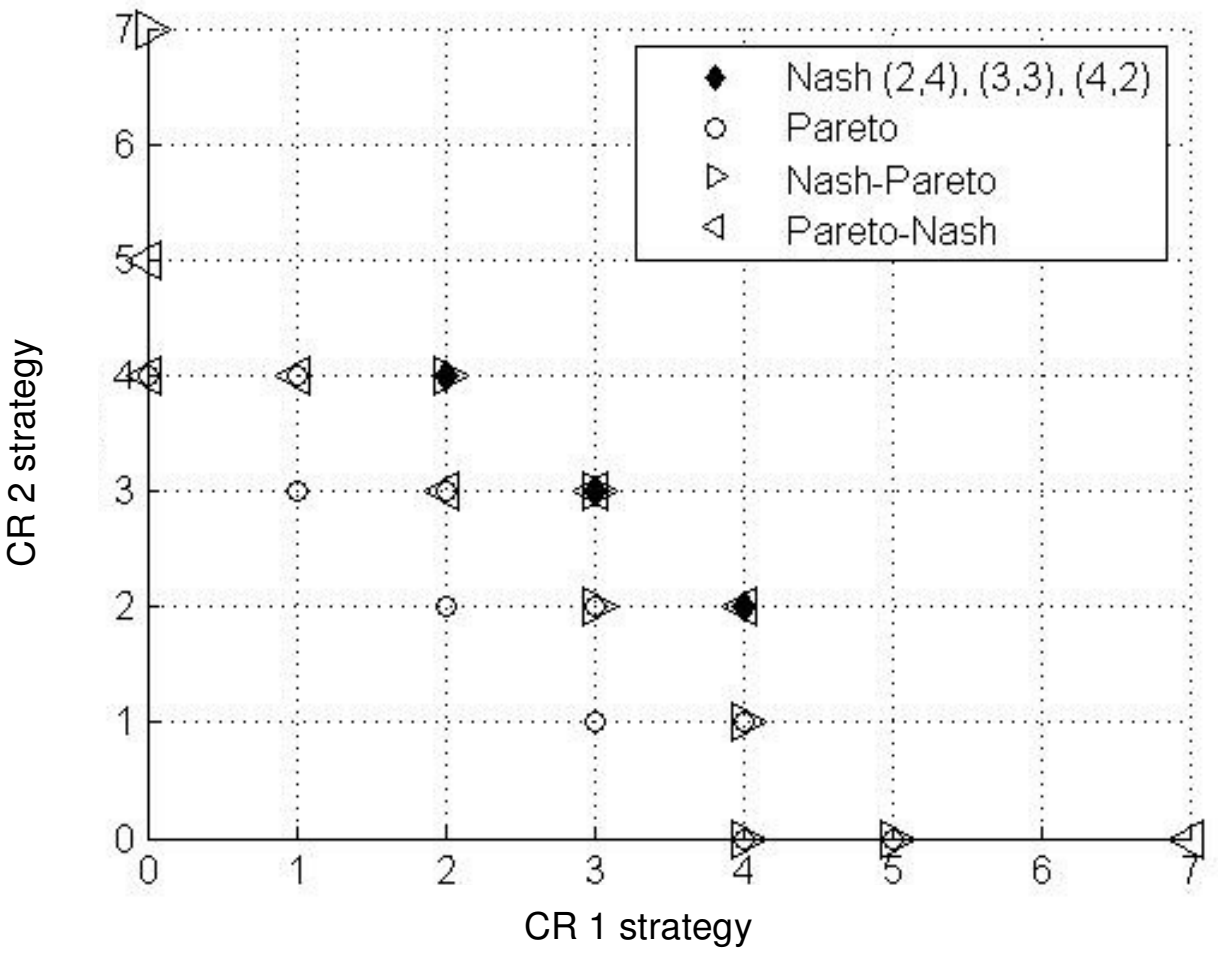

Figure 3. Discrete Cournot modelling – two radios (W=10, K=1). Evolutionary detected equilibrium strategies: Nash (2,4), (3,3), (4,2) Pareto, Nash-Pareto, and Pareto-Nash.

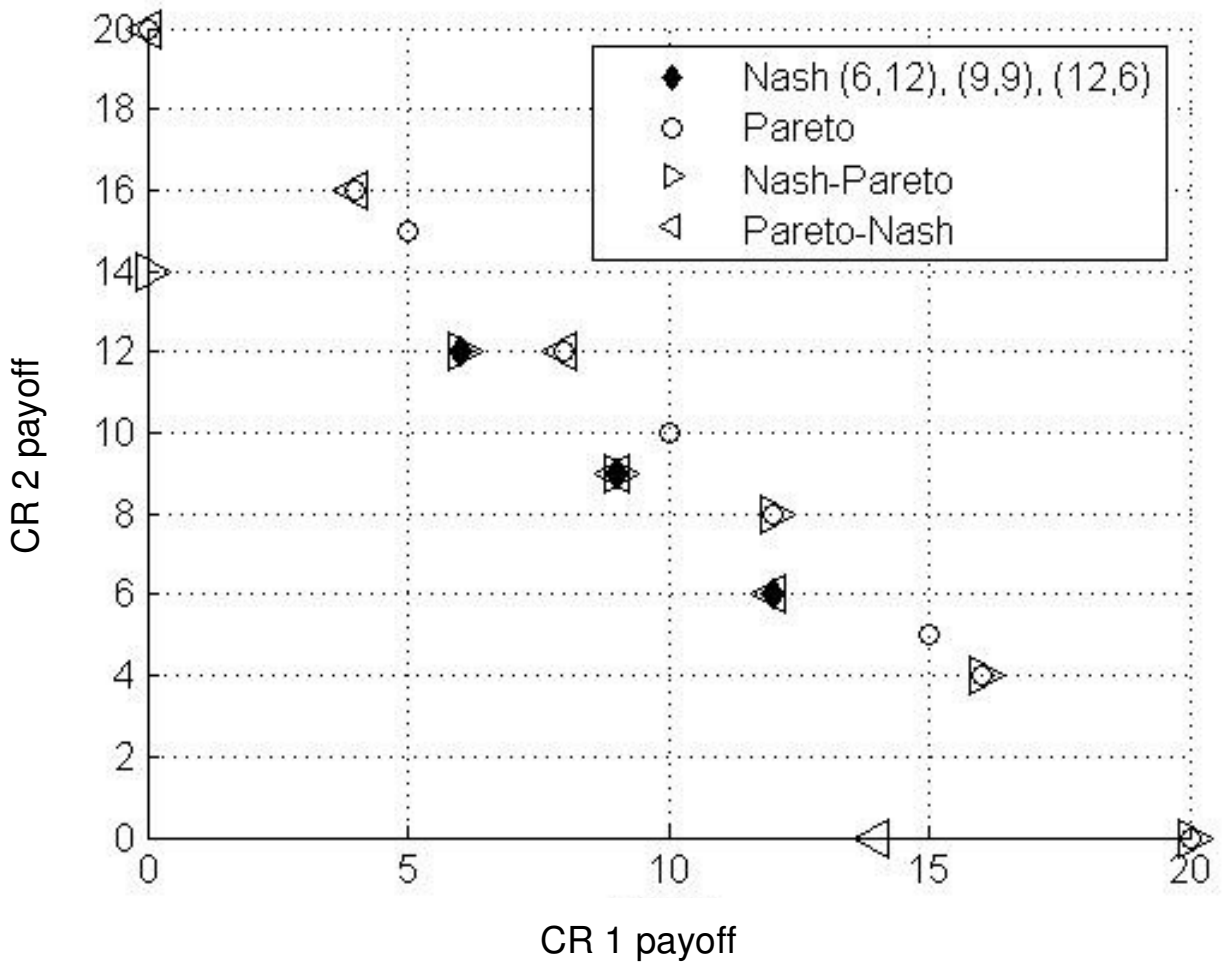

Figure 4. Discrete Cournot modelling – two radios (W=10, K=1). Payoffs of the evolutionary detected equilibria: Nash (6,12), (9,9), (12,6), Pareto, N-P, and P-N.

As expected, when 3 CRs share the same set of available channels, the number of accessed channels per CR and their respective payoffs decrease (compared to the 2-CR access scenario): NE strategy is (2.25, 2.25, 2.25) and NE payoff is (5.05, 5.05, 5.05).

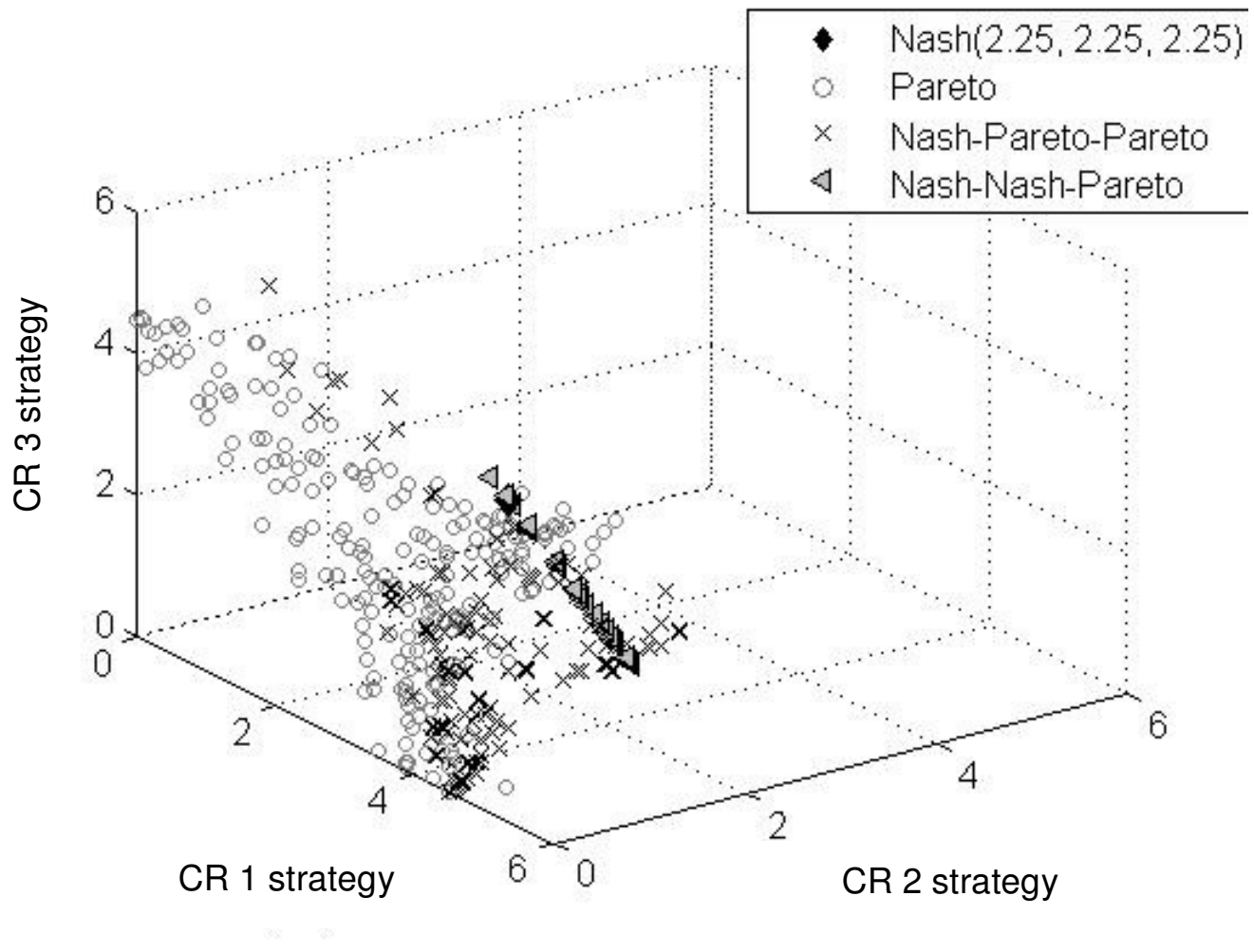

Figure 5. Continuous Cournot modelling – three radios (W=10, K=1). Evolutionary detected equilibrium strategies: Nash (2.25, 2.25, 2.25), Pareto, N-N-P, and N-P-P.

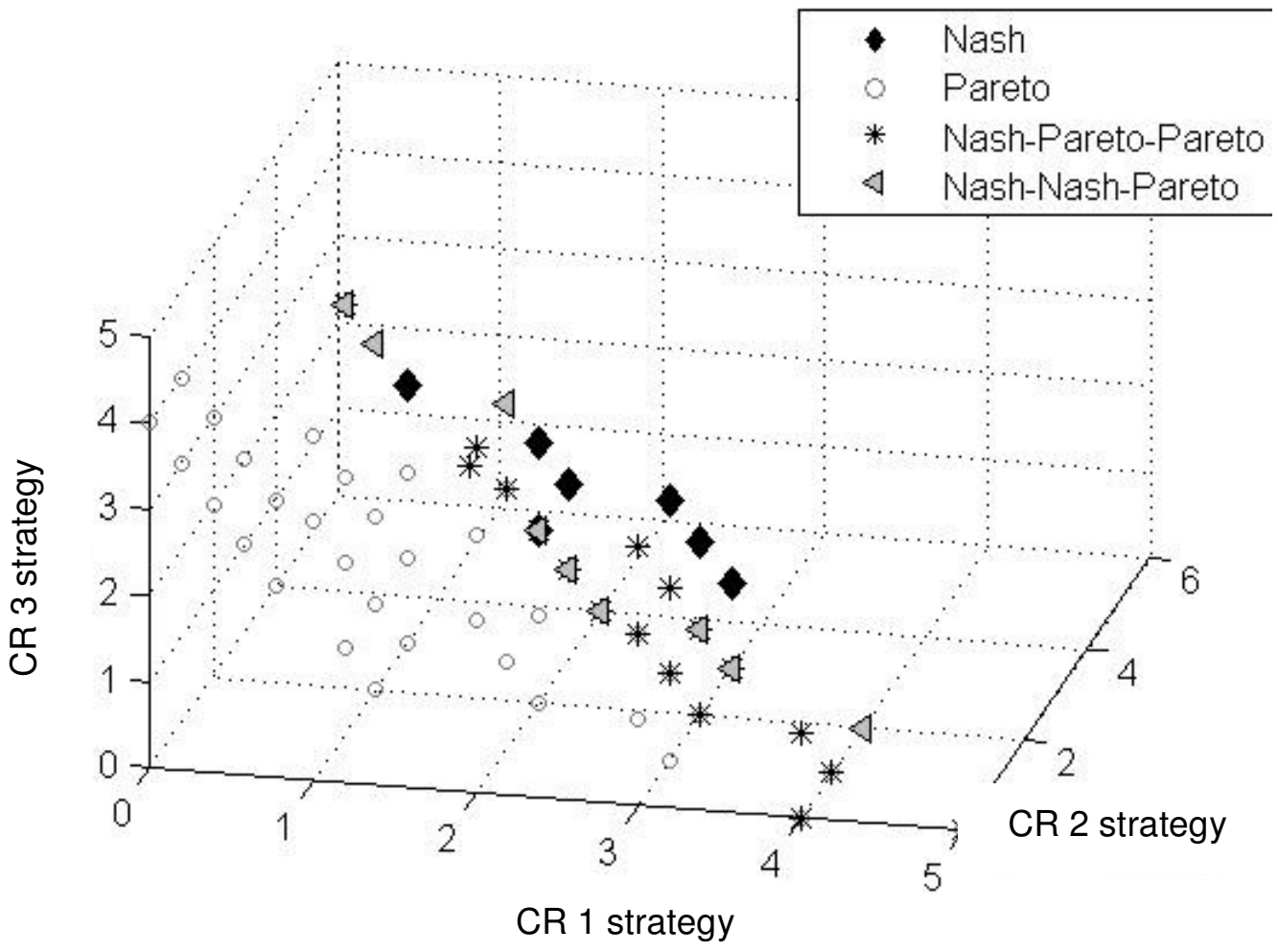

Figure 7. Discrete Cournot modelling, 3 CRs (W=10, K=1). Strategies: Nash: (2,2,2), (2,2,3), (2,3,2), (3,2,2), (1,3,3), (3,1,3), (3,3,1), Pareto, N-N-P, and N-P-P.

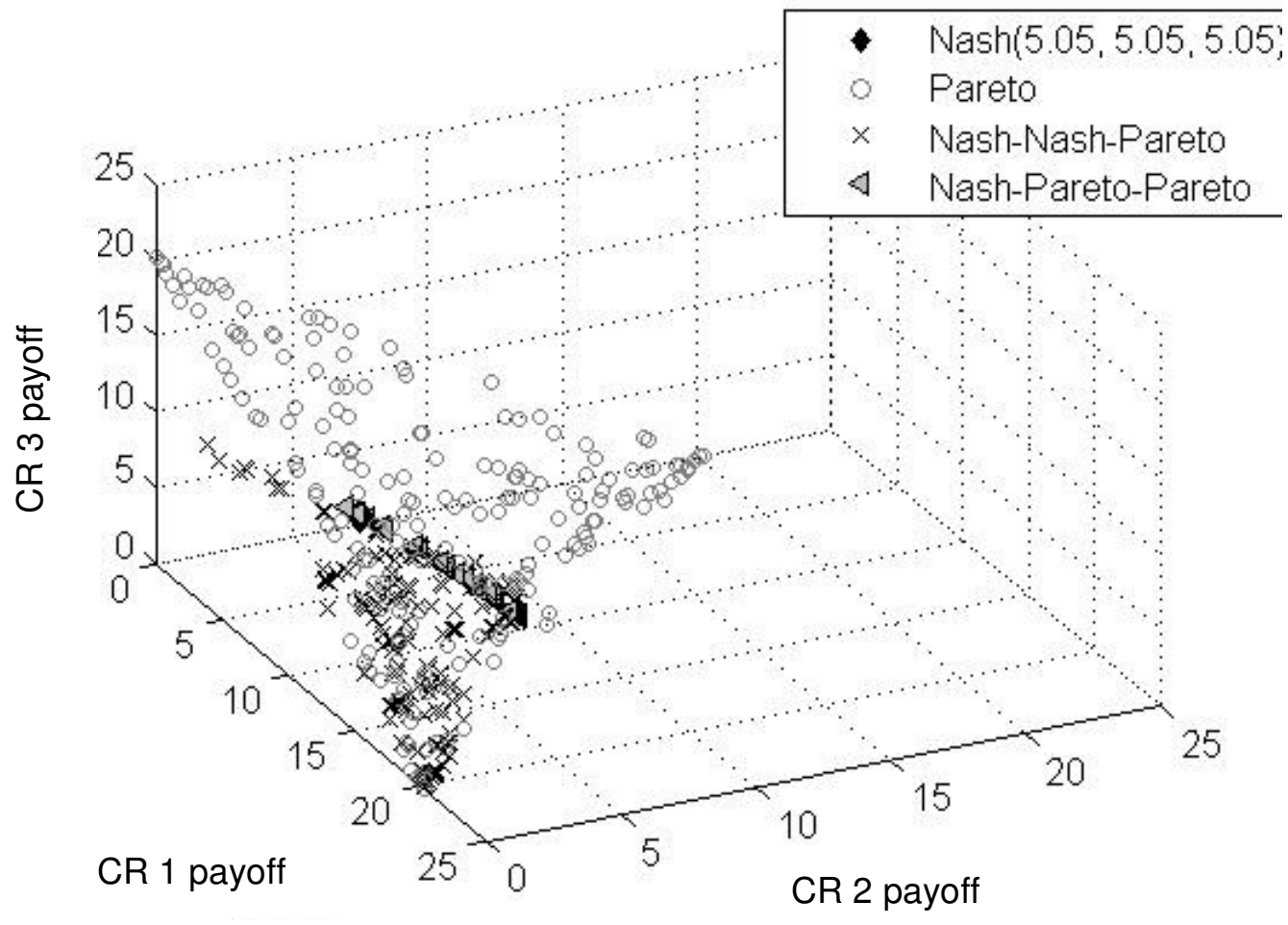

Figure 6. Continuous Cournot modelling – 3 radios (W=10, K=1). Payoffs of the evolutionary detected equilibria: Nash (5.05, 5.05, 5.05), Pareto, N-N-P, and N-P-P.

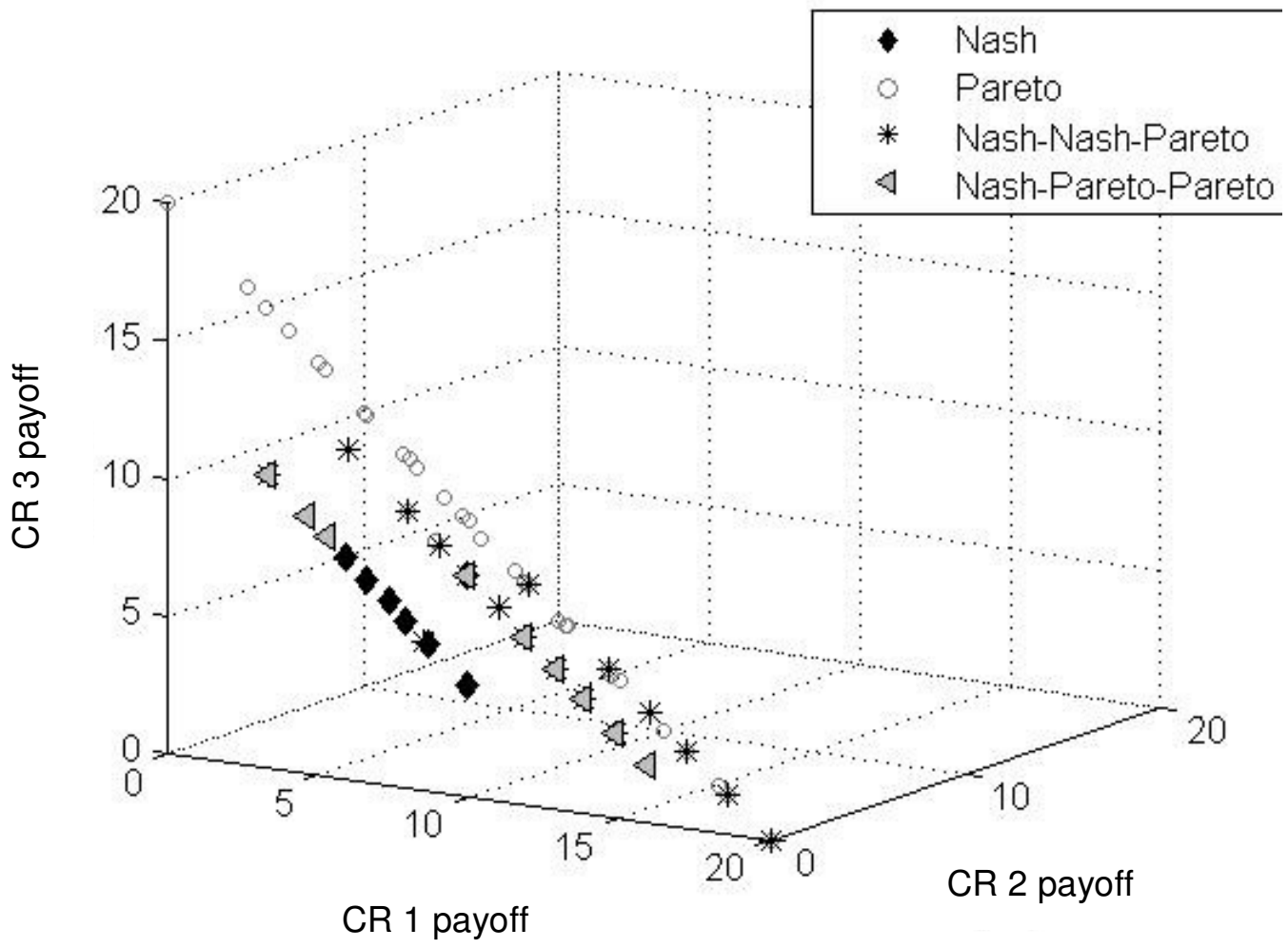

Figure 8. Discrete Cournot modelling, 3 CRs (W=10, K=1). Payoffs: Nash (6, 6, 6), (4, 4, 6), (4, 6, 4), (6, 4, 4), (2, 6, 6), (6, 2, 6), (6, 6, 2), Pareto, N-N-P, and N-P-P.

In the 3-player game the variety of joint Nash-Pareto equilibrium situations increases combinatorially. For illustration we chose Nash-Nash-Pareto and Pareto-Pareto-Nash equilibria (Fig. 5 and Fig. 6). In order to get a better view of the equilibria we turn to the discrete instance of the game.

### *D. Discrete Cournot modelling, 3-player simultaneous access*

Fig. 7 and Fig. 8 capture the equilibrium situations for the discrete modelling of the 3-CR simultaneous access.

Seven Nash equilibria (2,2,2), (2,2,3), (2,3,2), (3,2,2), (1,3,3), (3,1,3), (3,3,1) can be identified. This indicates an even higher flexibility in choosing the number of accessed channels for each CR. Also the range of satisfactory payoffs is increased. NE payoffs are (6, 6, 6), (4, 4, 6), (4, 6, 4) , (6, 4, 4), (2, 6, 6), (6, 2, 6), (6, 6, 2).

We may notice that N-N-P and P-P-N no longer overlap Pareto and Nash equilibria. The heterogeneity of players has a visible impact on the game equilibria – new equilibrium situations appear (the distinct Nash-Pareto equilibria).

## V. CONCLUSIONS

One-shot games are considered in order to assess non-iterative spectrum access scenarios. Open spectrum access scenarios are modelled based on continuous and discrete reformulations of the Cournot game theoretical model. Simultaneous access of two and three CRs is analyzed. Nash and Pareto equilibria are revisited in the discrete instance of the game. Heterogeneity of players is captured by joint Nash-Pareto equilibria.

Numerical simulations reveal equilibrium situations that may be reached in simultaneous access scenarios. Besides

Nash and Pareto equilibria new equilibrium situations establish, especially in the 3-player game.

Continuous Cournot modelling does not capture the variety of possible situations. Discrete instances of the game reveal multiple Nash equilibria and distinct Nash-Pareto equilibria. This indicates more flexibility for the CRs in choosing satisfactory spectrum access strategies.

## Acknowledgment

This paper was supported by CNCSIS –UEFISCDI, Romania, PD, project number 637/2010. This work was also supported by the project "Develop and support multidisciplinary postdoctoral programs in primordial technical areas of national strategy of the research - development - innovation" 4D-POSTDOC, contract nr. POSDRU/89/1.5/S/52603, project co-funded from European Social Fund through Sectorial Operational Program Human Resources 2007-2013. D. Dumitrescu acknowledges the support of a grant from the John Templeton Foundation. The opinions expressed in this publication are those of the authors and do not necessarily reflect the views of the John Templeton Foundation. R. Nagy acknowledges the support of POSDRU/88/1.5/S/60185 – "Innovative Doctoral Studies in a Knowledge Based Society".

## References

[1] I. Akyildiz, W. Lee, M. Vuran, S. Mohanty, "NeXt generation/dynamic spectrum access/cognitive radio wireless networks: a survey," *Computer Networks: The Int. J. of Computer and Telecommunications Networking*, vol. 50, no. 13, pp. 2127 –2159, 2006.

[2] C. Cordeiro, K. Challapali, D. Birru, "IEEE 802.22: An Introduction to the first wireless standard based on cognitive radios", *J. of Comm.*, vol. 1, no.1, pp.38-47, 2006.

[3] S. Deb, V. Srinivasan, R. Maheswari, "Dynamic Spectrum Access in DTV Whitespaces: Design Rules, Architecture and Algorithms", MobiCom2009.

[4] S. Srinivasa, S.A. Jafar, "The Throughput Potential of Cognitive Radio: A Theoretical Perspective", *IEEE Comm. Mag.*, pp.73-79, 2007.

[5] Linda E. Doyle, *Essentials of Cognitive Radio*, Cambridge U.P., 2009.

[6] P. Kolodzy, "Communications policy and spectrum management", in Fette, Bruce, A. (Ed.): *Cognitive Radio Technology*, Elsevier, 2006, 1st edn., pp. 29-72.

[7] D. Niyato, E. Hossain, "Microeconomic models for dynamic spectrum management in cognitive radio networks", *Cognitive Wireless Communication Networks*, Hossain, E., Bhargava, V.K. (eds.), Springer Science+Business Media, NY, pp. 391-423, 2007.

[8] M.J. Osborne, *An Introduction to Game Theory*, Oxford U.P., 2004.

[9] P. Weiser, D. Hatfield, "Policing the spectrum commons," *Fordham Law Review*, vol. 74, pp. 101–131, 2005.

[10] B. Wang, Y. Wu, K. J. R. Liu, "Game theory for cognitive radio networks: an overview", *Computer Networks*, *The Int. J. of Computer and Telecommunications Networking*, vol. 54, no. 14, pp.2537-2561, 2010.

[11] J. Nash, "Non-Cooperative Games", *The Annals of Mathematics,* vol. 54, no. 2, pp. 286-295, 1951.

[12] J.O. Neel, J.H. Reed, R.P. Gilles, "Game Models for Cognitive Radio Algorithm Analysis",
http://www.mprg.org/people/gametheory/presentations.shtml , 2004.

[13] A. MacKenzie, S. Wicker, "Game Theory in Communications: Motivation, Explanation, and Application to Power Control", Globecom2001, pp. 821-825, 2001.

[14] J.W. Huang, V. Krishnamurthy, "Game Theoretic Issues in Cognitive Radio Systems", *J. of Comm.*, vol.4, no.10, pp.790-802, 2009.

[15] D. Dumitrescu, R. I. Lung, T. D. Mihoc, "Evolutionary Equilibria Detection in Non-cooperative Games", EvoStar2009, Applications of Evolutionary Computing, *Lecture Notes in Computer Science*, Springer Berlin / Heidelberg, Vol. 5484, pp. 253-262, 2009.

[16] M. Maskery, V. Krishnamurthy, Q. Zhao, "GT Learning and Pricing for Dynamic Spectrum Access in Cognitive Radio", E. Hossain, V.K. Bhargava, (eds.) *Cognitive Wireless Communication Networks*, Springer Science +Business Media., NY, 2007.

[17] D. Monderer, L. Shapley, "Potential Games, Games and Economic Behavior", 14, pp. 124–143, 1996.

[18] D. Niyato, E. Hossain, "Microeconomic Models for Dynamic Spectrum Management in Cognitive Radio Networks", E. Hossain, V.K. Bhargava, (eds.) *Cognitive Wireless Communication Networks*, Springer Science+Business Media, NY, 2007.

[19] K. Deb, S. Agrawal, A. Pratab, T. Meyarivan, "A Fast Elitist Non-Dominated Sorting Genetic Algorithm for Multi-Objective Optimization: NSGA-II", Schoenauer, M., Deb, K., Rudolph, G., Yao, X., Lutton, E., Merelo, J. J., Schwefel, H.P. (eds.) Proc. of the PPSN2000, Springer *LNCS*, 1917, pp.849-858, 2000.

[20] Ligia C. Cremene, D. Dumitrescu, Réka Nagy, "Oligopoly Game Modeling for Cognitive Radio Environments", Balandin S., Dunaytsev, R., Koucheryavy, Y., (eds.) Proc. of the Smart Spaces and Next Gen. Wired/Wireless Networking, Springer *LNCS*, Vol 6294, pp. 219-230, 2010.

[21] R.I. Lung, D. Dumitrescu, "Computing Nash Equilibria by Means of Evolutionary Computation", *Int. J. Comput Commun*, Vol.3, pp:364-368, 2008.